\documentclass[twocolumn]{aastex631}

\begin{document}

\title{Discovery of a large-scale \ion{H}{1} plume in the NGC 7194 Group}


\author[0000-0002-0786-7307]{Mina Pak}
\affiliation{School of Mathematical and Physical Sciences, Macquarie University, NSW 2109, Australia}
\affiliation{ARC Centre of Excellence for All Sky Astropysics in 3 Dimensions (ASTRO 3D), Australia}

\author[0000-0002-3744-6714]{Junhyun Baek}
\affiliation{Department of Astronomy, Yonsei University, 50 Yonsei-ro, Seodaemun-gu, Seoul, 03722, Republic of Korea}

\author[0000-0003-3451-0925]{Joon Hyeop Lee}
\affiliation{Korea Astronomy and Space Science Institute (KASI), 776 Daeduk-daero, Yuseong-gu, Daejeon  34055, Republic of Korea}

\author[0000-0003-1440-8552]{Aeree Chung}
\affiliation{Department of Astronomy, Yonsei University, 50 Yonsei-ro, Seodaemun-gu, Seoul, 03722, Republic of Korea}

\author[0000-0002-2879-1663]{Matt Owers}
\affiliation{School of Mathematical and Physical Sciences, Macquarie University, NSW 2109, Australia}
\affiliation{ARC Centre of Excellence for All Sky Astropysics in 3 Dimensions (ASTRO 3D), Australia}

\author[0000-0002-0145-9556]{Hyunjin Jeong}
\affiliation{Korea Astronomy and Space Science Institute (KASI), 776 Daeduk-daero, Yuseong-gu, Daejeon  34055, Republic of Korea}

\author{Eon-Chang Sung}
\affiliation{Korea Astronomy and Space Science Institute (KASI), 776 Daeduk-daero, Yuseong-gu, Daejeon  34055, Republic of Korea}

\author[0000-0002-3211-9431]{Yun-Kyeong Sheen}
\affiliation{Korea Astronomy and Space Science Institute (KASI), 776 Daeduk-daero, Yuseong-gu, Daejeon  34055, Republic of Korea}

\begin{abstract}
We present the discovery of a new \ion{H}{1} structure in the NGC 7194 group from the observations using the Karl G. Jansky Very Large Array. NGC 7194 group is a nearby (z\,$\sim$\,0.027) small galaxy group with five quiescent members. The observations reveal a 200\,kpc-long \ion{H}{1} plume that spans the entire group with a total mass of M$_{HI} = 3.4 \times 10^{10}$\,M$_{\odot}$. The line-of-sight velocity of the \ion{H}{1} gas gradually increases from south (7200\,\,km\,s$^{-1}$) to north (8200\,km\,s$^{-1}$), and the local velocity dispersion is up to 70\,km\,s$^{-1}$. The structure is not spatially coincident with any member galaxies but it shows close associations with a number of blue star-forming knots. Intragroup \ion{H}{1} gas is not rare, but this particular structure is still one of the unusual cases in the sense that it does not show any clear connection with sizable galaxies in the group. We discuss the potential origins of this large-scale \ion{H}{1} gas in the NGC\,7194 group and its relation with the intergalactic star-forming knots. We propose that this HI feature could have originated from tidal interactions among group members or the infall of a late-type galaxy into the group. Alternatively, it might be leftover gas from flyby intruders.

\end{abstract}


\section{Introduction}

Cosmological simulations predict that galaxies and groups are assembled into massive clusters hierarchically. Thus it is likely that a considerable fraction of present-day cluster galaxies have experienced morphological transformation in galaxy groups, which allow more violent mergers between galaxies (pre-processing, \citealt{Zab98}; \citealt{DeL12}). Recent observational and theoretical studies emphasize the importance of galaxy groups as the environments for pre-processing (\citealt{Cor06}; \citealt{Jus19}; \citealt{Kle21}; \citealt{Sco22}). 

Neutral Hydrogen (\ion{H}{1}), a low column density diffuse atomic gas, is a great tool to probe gravitational and/or hydrodynamic interactions in galaxy groups (\citealt{Yun94}; \citealt{Kil06}; \citealt{Ser13}; \citealt{Oos18}; \citealt{Sap18}; \citealt{Kle19}; \citealt{Kor20}). The \ion{H}{1} that resides in galaxies is typically a factor of 2-3 larger than the stellar component (\citealt{Pis00}; \citealt{Chu09}; \citealt{Kor18}), therefore it is more susceptible to external processes such as tidal and hydro-dynamical processes in galaxies and the intracluster medium (\citealt{Gun72}; \citealt{Cow77}; \citealt{Lar80}; \citealt{Nul82}; \citealt{Moo96}; \citealt{Chu07, Chu09}; \citealt{Ras08}). During interactions, \ion{H}{1} gas is scattered to circumgalactic regions such as isolated dark cloud \citep{Can15}, and then forms new stars and dwarf galaxies (\citealt{Thi09}; \citealt{Cor21}). Therefore, studying the properties of \ion{H}{1} in galaxy groups is important to understand the pre-processing.

The \ion{H}{1} observations in many galaxy groups have been widely used to study galaxy evolution and therefore the structures of \ion{H}{1} clouds, streams, plumes and tidal tails are commonly found in nearby groups. (e.g., \citealt{Kor03}; \citealt{Kor04}; \citealt{Eng10}; \citealt{LWa19}; \citealt{Ser19}; \citealt{Kle21}; \citealt{Nam21}). The triplet of M81 group is a well-known example of tidal interactions among the galaxies revealed by tidal \ion{H}{1} bridges and disturbed \ion{H}{1} distributions (\citealt{Got75}; \citealt{Yun94}). \citet{Ser13} reported a long \ion{H}{1} tail in HCG 44 group, which may be the result of an interaction between the group and a spiral galaxy. The \ion{H}{1} ring of the NGC 5291 galaxy is known to have formed through a violent collision with a neighbor \citep{Bou07}. The active star formation has been triggered, particularly in the form of giant HII regions, which also take place as the formation of dwarf galaxies (\citealt{Mal97}; \citealt{Duc98}; \citealt{Boq07}; \citealt{Fen16}). One of the most well-known examples is the giant \ion{H}{1} ring ($\sim$200\,kpc in diameter) in the Leo I group \citep{Sch83} but its origin still shrouded in mystery. The Leo ring may be a collisional ring \citep{Mic10} or the stripped remnants of tidal interactions between galaxies (\citealt{Bek05}; \citealt{Cor21}). The Leo ring is isolated, not associated with any luminous galaxy in the group, and is much more quiescent than many other collisional ring galaxies (\citealt{Hig95}; \citealt{Cor21}; \citealt{Fen16}). Meanwhile, ever since the discovery of \ion{H}{1} in even early-type galaxies (ETGs, \citealt{Gou69}), the idea that ETGs can harbor \ion{H}{1} gas has become well-known now, supported by numerous observational studies of nearby ETGs (\citealt{Oos10}; \citealt{Ser12}; \citealt{Bai20}.)

In this Letter, we report the discovery of an extremely extended \ion{H}{1} structure in the NGC\,7194 group from our the Karl G. Jansky Very Large Array (VLA) observations. The NGC\,7194 group is a nearby (z\,$\sim$\,0.027) galaxy group with velocity dispersion of $\sigma \sim$\,201\,km\,s$^{-1}$, mass of 2\,$\times$\,10$^{13}$\,M$_{\odot}$ and virial radius of 460\,kpc \citep{Tul15}. An interesting point is that four of the five members are obviously quiescent and one shows only a very weak sign of star formation, which are not expected to hold a large amount of \ion{H}{1} gas. Intriguingly, there are two distinct galaxies in transition: a passive spiral (PSp) galaxy and a shell galaxy, which may have recently experienced the transformation of morphologies by group environment effects. Thus, the NGC\,7194 group is an excellent target to investigate group environment processes acting on galaxies with the combination of optical and \ion{H}{1} observations. The \ion{H}{1} data from the Arecibo Legacy Fast ALFA Extragalactic \ion{H}{1} Source Catalog (ALFALFA; \citealt{Hay18}) give strong evidence of an amount of \ion{H}{1} in the NGC\,7194 group. Did the massive \ion{H}{1} gas originate from the passive members? Otherwise, where did the gas come from? Finding an answer to the origin of the \ion{H}{1} and those intergalactic features must be the key to understanding the current dynamical status of this environment and the evolution of galaxies in the group. Throughout the paper, we adopt a standard $\Lambda$CDM cosmology with $\Omega_m=0.3$, $\Omega_\Lambda=0.7$, and $H_0=67.8$\,km\,s$^{-1}$\,Mpc$^{-1}$. This gives a scale of 0.557\,kpc/\arcsec.

\section{The Karl G. Jansky Very Large Array Observations and Analysis} 
We observed the NGC\,7194 group with the VLA D-configuration in L-band (project id: 22A-161). The pointing center is the position of the NGC\,7194 at R.A.:\.22$^{\rm h}$\,03$^{\rm m}$\,30.$^{\rm s}$936, DEC.:\,+12\,$^{\circ}$\,38\,$\arcmin$\,12\,.$\arcsec$415 (J2000). Two observation runs performed on 2022 Aug. 29 and Sep. 04 were used. The total on-source time is 100\,min. A single band of 16\,MHz width was centered at 1.385\,GHz with 31.25\,kHz channel separation, corresponding to $\sim$6.8\,km\,s$^{-1}$ at $z$\,=\,0.027. 3C~48 was observed as flux and bandpass calibrator and J2148+0657 was observed as complex gain calibrator

The data were processed using the standard CASA (5.6.2-3) pipeline. We first manually flagged bad data in the time and frequency domain using {\tt tfcrop}. Then we ran the pipeline for flux, bandpass, and complex gain calibration. After the pipeline calibration of each dataset, we merge two epoch data for imaging using {\tt tclean}. In order to find \ion{H}{1} detect the \ion{H}{1} line, we subtracted the continuum in the image plane using line free channels by {\tt imcontsub}. In this Letter, Briggs weighting of a robust~1 in the CASA scale was adopted, yielding a synthesized beam of 58\,.$\arcsec$2 $\times$ 51\,.$\arcsec$7 (32.4 $\times$ 28.8\,kpc). The velocity of the spectral cube was smoothed to $\sim$20\,km\,s$^{-1}$ (3 channels) to maximise the signal-to-noise ratio (S/N) while keeping the spectral resolution good enough to study the kinematics of the \ion{H}{1} gas. The smoothed 20 km/s channel rms is 0.6 mJy/beam. \ion{H}{1} moment maps were extracted by applying a 5-channel Hanning smoothing filter in the spectral axis and a 4-pixel Gaussian smoothing filter in spatial axes, with a cutoff level of 1.5 times channel map $rms$. This allows us to probe the faint \ion{H}{1} structure in the outskirt.

\section{Extended \ion{H}{1} structure in the NGC 7194 group}
\begin{figure*}
\centering{\includegraphics[width=17.5cm]{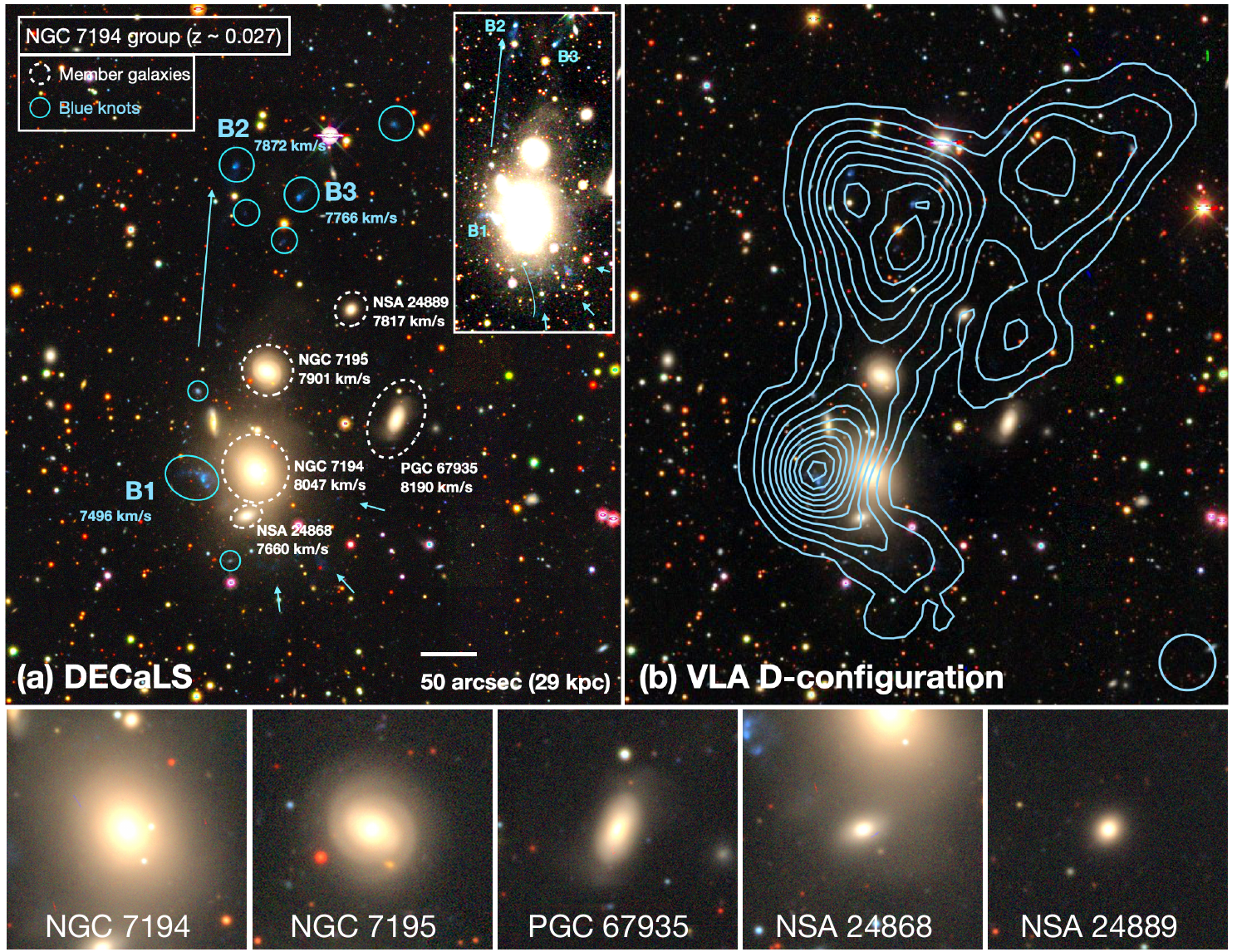}}
\caption{(a) The DECaLS image of the NGC 7194 group. North is at the top and east is to the left. White dotted circles are the five previously-known members of the group. Cyan circles are blue knots. The long arrow is drawn parallel with the long diffuse and blue stream, and the small arrows mark diffuse blue branches in the south-western outskirts of the N7194. (b) \ion{H}{1} distributions from VLA D-configuration is overlaid on the DECaLS image. Circle in the right-bottom corner denotes the beam size of D-configuration. The contour levels of \ion{H}{1} column density go from 1.5 to 47.0 $\times$ 10$^{19}$ cm$^{-2}$ in the interval with 10 steps in linear scale. The DECaLS images (70$\arcsec$\,$\times$\,70$\arcsec$) of five quiescent galaxies are shown in the bottom panel. The IDs of NSA 24868 and NSA 24889 are taken from the NASA-Sloan Atlas catalog (NSA, \citealt{Bla11}).}
\label{F1}
\end{figure*}
\begin{figure*}
\centering{\includegraphics[width=18cm]{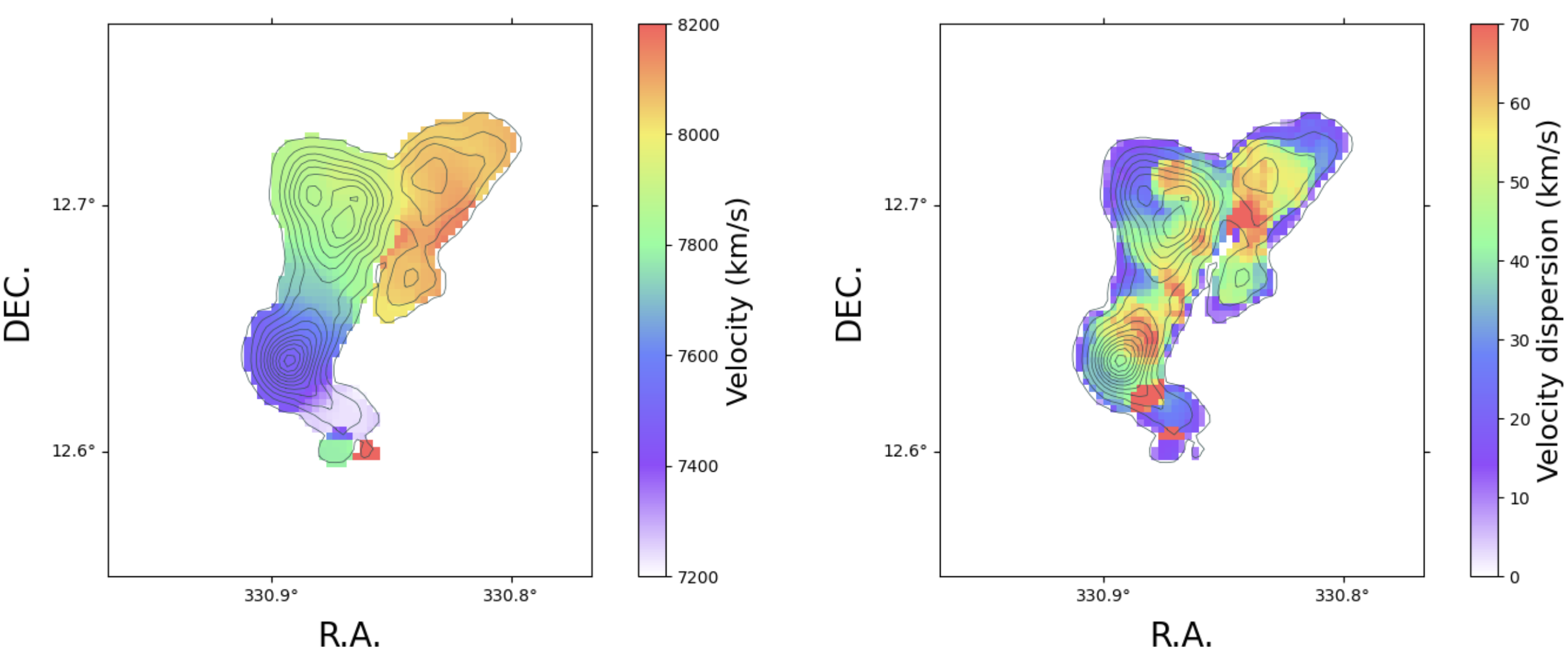}}
\caption{The \ion{H}{1} intensity weighted velocity (left panel) and velocity dispersion (right panel) from VLA D-configuration. \ion{H}{1} column density contours are overlaid with solid lines on both maps. The \ion{H}{1} velocities increase smoothly from 7200\,km\,s$^{-1}$ in the south to 8200\,km\,s$^{-1}$ in the north. }
\label{F2}
\end{figure*}

We present the \ion{H}{1} structure from the VLA observations overlaid on the optical image from DECam Legacy Survey (DECaLS) in Figure \ref{F1}. The NGC\,7194 group consists of five confirmed member galaxies \citep{Tul15} and none of them form stars actively with no strong nebular emissions in their optical spectroscopy. Aside from the passive member galaxies, Figure \ref{F1}a reveals several blue star-forming knots and a faint, long, linear stellar stream. These features may indicate evidence of gravitational interactions. B1, close to the brightest group galaxy (NGC 7194), is the biggest star-forming knots in this group. A long arrow from B1 to B2 (and B3) marks the faint stellar stream with a projected distance of 160\,kpc, which may be a connecting bridge between B1 and B2 (and B3). There are blue and diffuse branches in the south-western outskirts of the NGC 7194 and the largest one extends to a projected distance of 45\,kpc (see the inset of Figure \ref{F1}). 

The \ion{H}{1} distribution in Figure \ref{F1}b reveals that the structure extends across the group. It is highly interesting that the largest blue knot B1 and the peak surface density of \ion{H}{1} gas almost overlap in position. It is also noticeable that all the blue knots are found inside the \ion{H}{1} gas. In addition, the diffuse branches in the south-western outskirts of the NGC\,7194 also seems to be associated with the extended \ion{H}{1} structure. 

The \ion{H}{1} gas velocity field in the left panel of Figure \ref{F2} reveals a gradient with velocities increasing from 7200\,km\,s$^{-1}$ in the south to 8200\,km\,s$^{-1}$ in the north. In the right panel of Figure \ref{F2}, the \ion{H}{1} velocity dispersion arise up to ($\sim$\,70\,km\,s$^{-1}$). While the intensity weighted velocities (left panel of Figure \ref{F2}) increase gradually from south to west, there are locally peaked high velocity dispersion regions ($\sim$70\,km\,s$^{-1}$) between bright \ion{H}{1} clumps, suggesting the presence of small kinematic discontinuities between bright \ion{H}{1} clumps.

The extent of the total \ion{H}{1} is 200\,kpc with \ion{H}{1} flux $= 9.85$\,Jy\,km\,s$^{-1}$ and M$_{HI} = 3.4 \times 10^{10}$\,M$_{\odot}$. NGC\,7194 group was covered by the Arecibo Legacy Fast Arecibo L-band Feed Array (ALFALFA) survey \citep{Hay18}. From the ALFALFA source catalogue from \citet{Hay18}, there are two detections of AGC323359 centered on R.A.:\,22$^{\rm h}$\,03$^{\rm m}$\,34.$^{\rm s}$6, DEC.:\,+12\,$^{\circ}$\,37\,$\arcmin$\,55\,.$\arcsec$ (J2000) which is near the B1, and AGC320009 centered on R.A.:\,22$^{\rm h}$\,03$^{\rm m}$\,31.$^{\rm s}$7, DEC.:\,+12\,$^{\circ}$\,39\,$\arcmin$\,28$\arcsec$ (J2000), which is NGC 7195, \ion{H}{1} flux densities are 2.03 and 5.28 Jy\,km\,s$^{-1}$, respectively (Haynes et al. 2018). However, our VLA probed \ion{H}{1} structure in the NGC\,7194 group is more extended than the ALFALFA field of view (FoV $\sim$3.5'). We detect 2.3 times more \ion{H}{1} flux densities than the ALFALFA reported \citep{Hay18}.

The densest \ion{H}{1} clump embracing B1 at $\sim$\,7480\,km\,s$^{-1}$ has \ion{H}{1} flux = 2.51\,Jy\,km\,s$^{-1}$ , M$_{HI}$ = 8.7 $\times 10^9$\,M$_{\odot}$. The upper area including B2 and B3 is more massive with \ion{H}{1} flux = 4.34\,Jy\,km\,s$^{-1}$, M$_{HI}$ = 1.5 $\times$ 10$^{10}$\,M$_{\odot}$. The \ion{H}{1} is extended further to the north-west without obvious optical counterparts.

\section{Discussion and Conclusion}

It is common to detect \ion{H}{1} gas in nearby galaxy groups, but the \ion{H}{1} structure in the NGC 7194 group is different from those previously known one, because those groups mostly consist of late-type and/or dusty galaxies (\citealt{Eng10}; \citealt{LWa19}; \citealt{Nam21}). Moreover, \ion{H}{1} gas is distributed on a scale of galaxies and the donor of \ion{H}{1} gas is rather obvious (\citealt{Mal97}; \citealt{Oos18}; \citealt{Bai20}). Based on the analysis currently available in this letter, we discuss several possible origins of the huge \ion{H}{1} structure in the NGC\,7194 group.

One plausible scenario is that the \ion{H}{1} gas originates from the member galaxies. For several decades, it has been reported that a significant number of early-type galaxies do indeed harbor \ion{H}{1} gas (\citealt{vanG97}; \citealt{Mor06}; \citealt{Ger16}; \citealt{Oos10}; \citealt{Ser12}). We confirmed that there is no strong sign of star formation for members by inspecting spectroscopy data using the Calar Alto Legacy Integral Field Area survey (NGC\,7194), the Sloan Digital Sky Survey (NGC 7195, PGC\,67935 and NSA\,24868), and Mapping Nearby Galaxies at APO (NSA\,24489) and photometry data using Galaxy Evolution Explorer (GALEX) and Wide-field Infrared Survey Explorer (WISE). The IDs of NSA 24868 and NSA 24889 are taken from the NASA-Sloan Atlas catalog (NSA, \citealt{Bla11}). PGC\,67935 is optically red but has a weak H$\alpha$ emission line at the center in the SDSS spectrum, which is a weak star-forming galaxy from the BPT diagram \citep{Bal81}. Its WISE color of W2\,-\,W3 is 2.19, which lies in the boundary between non star forming and weak star-forming galaxies. There is no detection in FUV. While all five members appear quiescent (see bottom panels for Figure \ref{F1}), several pieces of evidence indicate they may have recently experienced morphological transformation. 

Even though the existence of \ion{H}{1} gas in a PSp galaxy has not been reported to date, NGC\,7195 might be the most prime candidate. NGC\,7195 is a typical PSp galaxy having no sign of significant star formation and distinct spiral arms. From simulations and observations, spiral arm structures are found to fade over several Gyrs, after gas is stripped (\citealt{Bek02}; \citealt{Pak19}; \citealt{Pak21}). Since the stellar stream is located at the north-eastern vicinity of the NGC\,7195, \ion{H}{1} gas may be stripped from the NGC\,7195 by interaction with the NGC\,7194 galaxy and/or other members. For instance, PGC\,67935 also has a clear shell structure and a diffuse stellar stream in the east, which is an evidence of merging or interactions. Unfortunately, the observed GALEX field is too shallow (exposure time of \,$\sim$\,192s) to detect faint features between members. The expected \ion{H}{1} mass range of NGC\,7195 is from 4.3\,$\times$\,10$^8$\,M$_{\odot}$ to 2.1\,$\times$\,10$^{10}$\,M$_{\odot}$ at the stellar mass of log(M$_{\star}$/M$_{\odot}$) = 10.53 \citep{Cat18}. However, even if we consider the upper limit of the expected \ion{H}{1} mass, this amount is smaller than the total \ion{H}{1} mass that we found in this group.

Secondly, we suspect that a gas-rich late-type progenitor may have fallen into the group from north to south of the group and it ends up B1 now. During infall, loosely bound \ion{H}{1} gas of the progenitor is stripped by the gravity of the group and/or the group members, and left behind along the trajectory of the progenitor, extending up to 200\,kpc. In the tails, blue knots are formed where the \ion{H}{1} gas is condensed enough to form new stars. Since there was no available spectroscopic data for these blue knots to check their group membership, we confirmed it using long-slit spectroscopy for B1, B2 and B3 with the Doyak $1.8-$m telescope of Bohyunsan Optical Astronomy Observatory in Korea. From these observations, we detected strong H$\alpha$ emission centered at 7496, 7872 and 7766\,km\,s$^{-1}$ for B1, B2 and B3, respectively, which are in good agreement with the \ion{H}{1} velocity distribution (Figure 3). 

In order to test the feasibility of this scenario, we estimate the crossing time between the progenitor, B1, and the nearest H$\alpha$ knot, B2. Assuming a typical galaxy speed in groups ($\sim$300\,km\,s$^{-1}$; \citealt{Smi07}; \citealt{Pif14}), the crossing time is at least $\sim$520\,Myr to move the projected distance of 160\,kpc from B2 to B1. This timescale is much longer than the star formation timescale traced by H$\alpha$ emission which is $\sim$10\,Myr \citep{Ler12}. Therefore, it is less likely that a gas-rich late-type infaller is a donor of HI gas in the group if the star formation in B2 occurs instantly from stripped gas.

In addition to the timescales, this scenario does not explain detailed features of the \ion{H}{1} structure. If the progenitor has fallen simply from north to south, it is hard to explain how the blue streams exist at the south-western outskirts of the NGC 7194 galaxy and why the \ion{H}{1} mass is greater in the northern region around B2 and B3. In addition, B1 is too small and amorphous to be a progenitor. The stellar mass of the B1 approximated from its SDSS luminosity and color is [M$_{\star}$/M$_{\odot}$]\,$\sim$\,$10^{7.5}$, which is within the stellar mass range of that expected of a dwarf galaxy. Though low mass late-type galaxies can hold an amount of \ion{H}{1} mass up to [M$_{HI}$/M$_{\odot}$]\,$\sim$\,$10^9$ \citep{Hun19}, this is thirty times smaller than the \ion{H}{1} mass found in the group.

It is also possible that a gas-rich progenitor has flown by and interacted with the group. One example is a long \ion{H}{1} tail of $\sim$\,300\,kpc in HCG\,44 group \citep{Ser13}. \citet{Ser13} suggested that the tail may be the result of an interaction between the group and a spiral galaxy, NGC\,3162, separated by 650\,kpc. According to their hypotheses, the NGC\,3162 passed through the group at high velocity and its gas was stripped in its trajectory. The lopsided morphology in the optical image, indicating recently perturbed via interactions, may support their hypotheses.

There is a candidate, PGC\,67927, which is located in a projected distance of $\sim$\,600\,kpc apart from the NGC\,7194 to the north-west. PGC\,67927 is a late-type galaxy with stellar mass log(M$_{\star}$/M$_{\odot}$)\,$\sim$\,10.5 from the NSA catalog. The morphology of PGC\,67927 in optical image shows asymmetry in its spiral arm and quenching in the bulge. PGC\,67927 contains \ion{H}{1} flux of S$_{HI}$\,=\,1.36\,Jy\,km\,s$^{-1}$ \citep{Spr05}. Assuming a distance of 118\,Mpc from the NASA/IPAC Extragalactic Database, this corresponds to the \ion{H}{1} mass of 4.5 $\times$ 10$^{9}$\,M$_{\odot}$ using the equation (2) in \citet{Chu09}, which means PGC\,67927 is already a gas-rich system at its stellar mass based on \ion{H}{1} scaling relation from \citet{Cat18}. Therefore, it is unlikely that PGC\,67927 left a significant amount of \ion{H}{1} over the NGC\,7194 group.

In conclusion, we have discovered a huge \ion{H}{1} structure in the small NGC\,7194 galaxy group. This isolated \ion{H}{1} cloud is extremely rare especially in a group environment. Our new discovery in this letter highlights that the origin of the structure is still shrouded in mystery, but evokes that the pre-processing in group environments can be very violent. In a forthcoming paper (Baek et al. in preparation), deeper insights on the detailed \ion{H}{1} properties of structures will come from the higher-resolution VLA observations in B- and C-configurations. We also will investigate the ionized gas properties of star-forming knots in optical data, which will allow us to probe their formation histories.

\begin{figure}
\centering{\includegraphics[width=7.5cm]{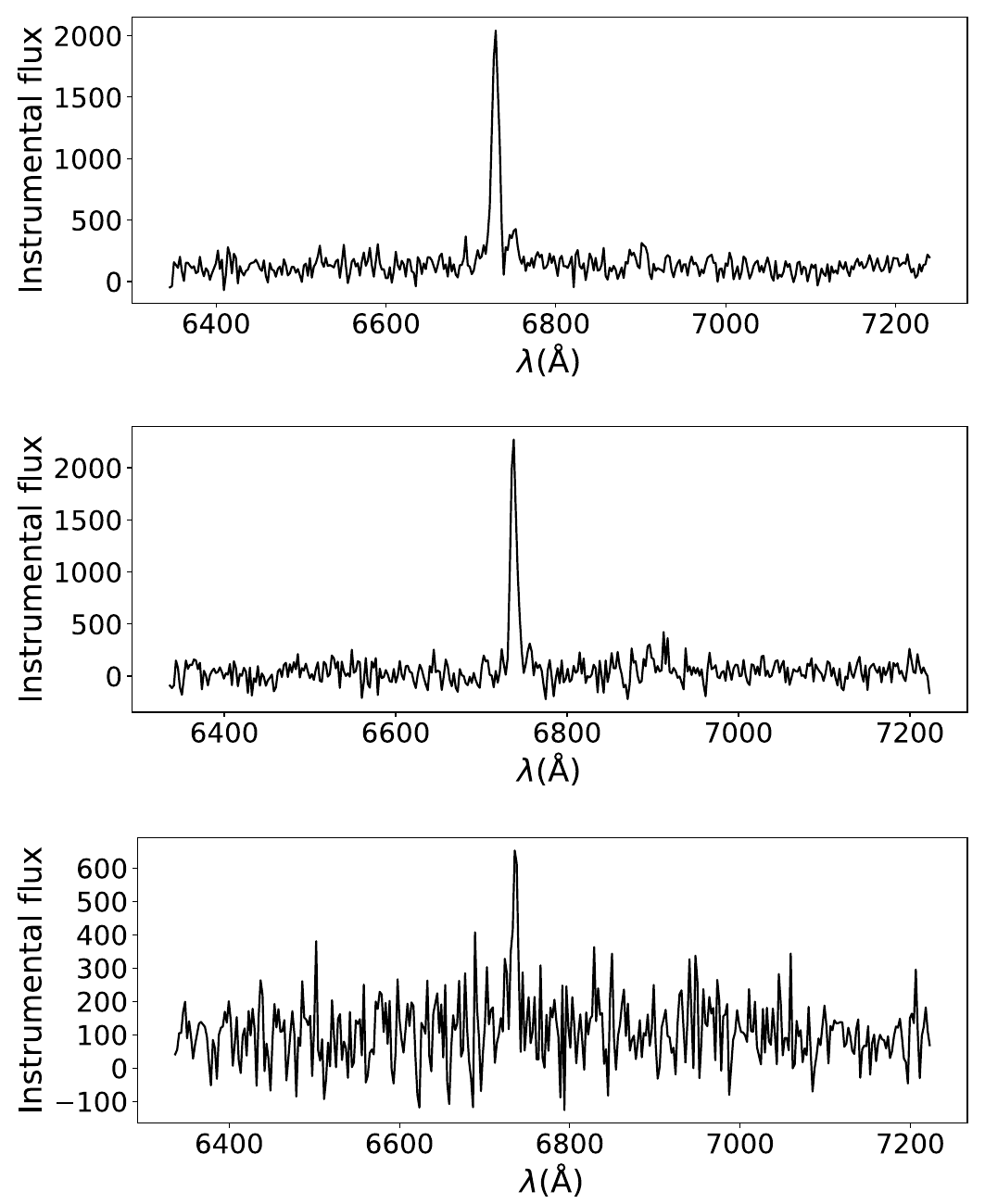}}
\caption{The spectra of three blue knots (B1 - B3) observed from the Doyak 1.8-m telescope of Bohyunsan Optical Astronomy Observatory long-slit spectrograph. We detected H$\alpha$ emission centered at 7496, 7872, and 7766 km\,s$^{-1}$, respectively. Instrumental flux is shown since calibration is not applied since signal to noise of the continuum is not sufficient enough to analyze flux calibration.}
\label{F3}
\end{figure}

\begin{acknowledgments}
We gratefully thank the anonymous referee for constructive comments that have significantly improved this manuscript.
We are grateful to Lachlan Marnoch for helpful discussions.
The National Radio Astronomy Observatory is a facility of the National Science Foundation operated under cooperative agreement by Associated Universities, Inc.
We thank the developers of the Bohyunsan Observatory and all staff of the Bohyunsan Optical Astronomy Observatory (BOAO).
This research was partially supported by the Australian Research Council Centre of Excellence for All Sky Astrophysics in 3 Dimensions (ASTRO 3D), through project number CE170100013.
J.B. and A.C. acknowledge support by the National Research Foundation of Korea (NRF), grant No. 2022R1A2C100298212, and No. 2022R1A6A1A03053472. This work was also supported by National R\&D Program through the NRF funded by the Korea government (Ministry of Science and ICT) (RS-2022-00197685).
J.H.L., H.J. and Y.K.S. acknowledge support from the NRF grants funded by the Korea government (MSIT) (No. 2022R1A2C1004025, No. NRF-2019R1F1A1041086 and No. 2019R1C1C1010279, respectively). 
This research was supported by the Korea Astronomy and Space Science Institute under the R\&D program (Projects No. 2023-1-830-01) supervised by the Ministry of Science and ICT. 
\end{acknowledgments}


\begin{thebibliography}{}
\expandafter\ifx\csname natexlab\endcsname\relax\def\natexlab#1{#1}\fi
\providecommand{\url}[1]{\href{#1}{#1}}
\providecommand{\dodoi}[1]{doi:~\href{http://doi.org/#1}{\nolinkurl{#1}}}
\providecommand{\doeprint}[1]{\href{http://ascl.net/#1}{\nolinkurl{http://ascl.net/#1}}}
\providecommand{\doarXiv}[1]{\href{https://arxiv.org/abs/#1}{\nolinkurl{https://arxiv.org/abs/#1}}}

\bibitem[{{Bait} {et~al.}(2020){Bait}, {Kurapati}, {Duc}, {Cuillandre}, {Wadadekar}, {Kamphuis}, \& {Barway}}]{Bai20}
{Bait}, O., {Kurapati}, S., {Duc}, P.-A., {et~al.} 2020, \mnras, 492, 1, \dodoi{10.1093/mnras/stz2972}

\bibitem[{{Baldwin} {et~al.}(1981){Baldwin}, {Phillips}, \& {Terlevich}}]{Bal81}
{Baldwin}, J.~A., {Phillips}, M.~M., \& {Terlevich}, R. 1981, \pasp, 93, 5, \dodoi{10.1086/130766}

\bibitem[{{Bekki} {et~al.}(2002){Bekki}, {Couch}, \& {Shioya}}]{Bek02}
{Bekki}, K., {Couch}, W.~J., \& {Shioya}, Y. 2002, \apj, 577, 651, \dodoi{10.1086/342221}

\bibitem[{{Bekki} {et~al.}(2005){Bekki}, {Koribalski}, {Ryder}, \& {Couch}}]{Bek05}
{Bekki}, K., {Koribalski}, B.~S., {Ryder}, S.~D., \& {Couch}, W.~J. 2005, \mnras, 357, L21, \dodoi{10.1111/j.1745-3933.2005.08625.x}

\bibitem[{{Blanton} {et~al.}(2011){Blanton}, {Kazin}, {Muna}, {Weaver}, \& {Price-Whelan}}]{Bla11}
{Blanton}, M.~R., {Kazin}, E., {Muna}, D., {Weaver}, B.~A., \& {Price-Whelan}, A. 2011, \aj, 142, 31, \dodoi{10.1088/0004-6256/142/1/31}

\bibitem[{{Boquien} {et~al.}(2007){Boquien}, {Duc}, {Braine}, {Brinks}, {Lisenfeld}, \& {Charmandaris}}]{Boq07}
{Boquien}, M., {Duc}, P.~A., {Braine}, J., {et~al.} 2007, \aap, 467, 93, \dodoi{10.1051/0004-6361:20066692}

\bibitem[{{Bournaud} {et~al.}(2007){Bournaud}, {Duc}, {Brinks}, {Boquien}, {Amram}, {Lisenfeld}, {Koribalski}, {Walter}, \& {Charmandaris}}]{Bou07}
{Bournaud}, F., {Duc}, P.-A., {Brinks}, E., {et~al.} 2007, Science, 316, 1166, \dodoi{10.1126/science.1142114}

\bibitem[{{Cannon} {et~al.}(2015){Cannon}, {Martinkus}, {Leisman}, {Haynes}, {Adams}, {Giovanelli}, {Hallenbeck}, {Janowiecki}, {Jones}, {J{\'o}zsa}, {Koopmann}, {Nichols}, {Papastergis}, {Rhode}, {Salzer}, \& {Troischt}}]{Can15}
{Cannon}, J.~M., {Martinkus}, C.~P., {Leisman}, L., {et~al.} 2015, \aj, 149, 72, \dodoi{10.1088/0004-6256/149/2/72}

\bibitem[{{Catinella} {et~al.}(2018){Catinella}, {Saintonge}, {Janowiecki}, {Cortese}, {Dav{\'e}}, {Lemonias}, {Cooper}, {Schiminovich}, {Hummels}, {Fabello}, {Ger{\'e}b}, {Kilborn}, \& {Wang}}]{Cat18}
{Catinella}, B., {Saintonge}, A., {Janowiecki}, S., {et~al.} 2018, \mnras, 476, 875, \dodoi{10.1093/mnras/sty089}

\bibitem[{{Chung} {et~al.}(2009){Chung}, {van Gorkom}, {Kenney}, {Crowl}, \& {Vollmer}}]{Chu09}
{Chung}, A., {van Gorkom}, J.~H., {Kenney}, J. D.~P., {Crowl}, H., \& {Vollmer}, B. 2009, \aj, 138, 1741, \dodoi{10.1088/0004-6256/138/6/1741}

\bibitem[{{Chung} {et~al.}(2007){Chung}, {van Gorkom}, {Kenney}, \& {Vollmer}}]{Chu07}
{Chung}, A., {van Gorkom}, J.~H., {Kenney}, J. D.~P., \& {Vollmer}, B. 2007, \apjl, 659, L115, \dodoi{10.1086/518034}

\bibitem[{{Comrie} {et~al.}(2021){Comrie}, {Wang}, {Hsu}, {Moraghan}, {Harris}, {Pang}, {Pi{\'n}ska}, {Chiang}, {Chang}, {Hwang}, {Jan}, {Lin}, \& {Simmonds}}]{Com21}
{Comrie}, A., {Wang}, K.-S., {Hsu}, S.-C., {et~al.} 2021, {CARTA: The Cube Analysis and Rendering Tool for Astronomy}, 2.0.0, Zenodo,  Zenodo, \dodoi{10.5281/zenodo.4905459}

\bibitem[{{Corbelli} {et~al.}(2021){Corbelli}, {Mannucci}, {Thilker}, {Cresci}, \& {Venturi}}]{Cor21}
{Corbelli}, E., {Mannucci}, F., {Thilker}, D., {Cresci}, G., \& {Venturi}, G. 2021, \aap, 651, A77, \dodoi{10.1051/0004-6361/202140398}

\bibitem[{{Cortese} {et~al.}(2006){Cortese}, {Gavazzi}, {Boselli}, {Franzetti}, {Kennicutt}, {O'Neil}, \& {Sakai}}]{Cor06}
{Cortese}, L., {Gavazzi}, G., {Boselli}, A., {et~al.} 2006, \aap, 453, 847, \dodoi{10.1051/0004-6361:20064873}

\bibitem[{{Cowie} \& {Songaila}(1977)}]{Cow77}
{Cowie}, L.~L., \& {Songaila}, A. 1977, \nat, 266, 501, \dodoi{10.1038/266501a0}

\bibitem[{{De Lucia} {et~al.}(2012){De Lucia}, {Weinmann}, {Poggianti}, {Arag{\'o}n-Salamanca}, \& {Zaritsky}}]{DeL12}
{De Lucia}, G., {Weinmann}, S., {Poggianti}, B.~M., {Arag{\'o}n-Salamanca}, A., \& {Zaritsky}, D. 2012, \mnras, 423, 1277, \dodoi{10.1111/j.1365-2966.2012.20983.x}

\bibitem[{{Duc} \& {Mirabel}(1998)}]{Duc98}
{Duc}, P.~A., \& {Mirabel}, I.~F. 1998, \aap, 333, 813

\bibitem[{{English} {et~al.}(2010){English}, {Koribalski}, {Bland-Hawthorn}, {Freeman}, \& {McCain}}]{Eng10}
{English}, J., {Koribalski}, B., {Bland-Hawthorn}, J., {Freeman}, K.~C., \& {McCain}, C.~F. 2010, \aj, 139, 102, \dodoi{10.1088/0004-6256/139/1/102}

\bibitem[{{Fensch} {et~al.}(2016){Fensch}, {Duc}, {Weilbacher}, {Boquien}, \& {Zackrisson}}]{Fen16}
{Fensch}, J., {Duc}, P.~A., {Weilbacher}, P.~M., {Boquien}, M., \& {Zackrisson}, E. 2016, \aap, 585, A79, \dodoi{10.1051/0004-6361/201527141}

\bibitem[{{Ger{\'e}b} {et~al.}(2016){Ger{\'e}b}, {Catinella}, {Cortese}, {Bekki}, {Moran}, \& {Schiminovich}}]{Ger16}
{Ger{\'e}b}, K., {Catinella}, B., {Cortese}, L., {et~al.} 2016, \mnras, 462, 382, \dodoi{10.1093/mnras/stw1675}

\bibitem[{{Gottesman} \& {Weliachew}(1975)}]{Got75}
{Gottesman}, S.~T., \& {Weliachew}, L. 1975, \apj, 195, 23, \dodoi{10.1086/153301}

\bibitem[{{Gouguenheim}(1969)}]{Gou69}
{Gouguenheim}, L. 1969, \aap, 3, 281

\bibitem[{{Greisen}(2003)}]{Gre03}
{Greisen}, E.~W. 2003, in Astrophysics and Space Science Library, Vol. 285, Information Handling in Astronomy - Historical Vistas, ed. A.~{Heck}, 109, \dodoi{10.1007/0-306-48080-8_7}

\bibitem[{{Gunn} \& {Gott}(1972)}]{Gun72}
{Gunn}, J.~E., \& {Gott}, III, J.~R. 1972, \apj, 176, 1, \dodoi{10.1086/151605}

\bibitem[{{Haynes} {et~al.}(2018){Haynes}, {Giovanelli}, {Kent}, {Adams}, {Balonek}, {Craig}, {Fertig}, {Finn}, {Giovanardi}, {Hallenbeck}, {Hess}, {Hoffman}, {Huang}, {Jones}, {Koopmann}, {Kornreich}, {Leisman}, {Miller}, {Moorman}, {O'Connor}, {O'Donoghue}, {Papastergis}, {Troischt}, {Stark}, \& {Xiao}}]{Hay18}
{Haynes}, M.~P., {Giovanelli}, R., {Kent}, B.~R., {et~al.} 2018, \apj, 861, 49, \dodoi{10.3847/1538-4357/aac956}

\bibitem[{{Higdon}(1995)}]{Hig95}
{Higdon}, J.~L. 1995, \apj, 455, 524, \dodoi{10.1086/176602}

\bibitem[{{Hunter} {et~al.}(2019){Hunter}, {Elmegreen}, \& {Berger}}]{Hun19}
{Hunter}, D.~A., {Elmegreen}, B.~G., \& {Berger}, C.~L. 2019, \aj, 157, 241, \dodoi{10.3847/1538-3881/ab1e54}

\bibitem[{{Jedrzejewski}(1987)}]{Jed87}
{Jedrzejewski}, R.~I. 1987, \mnras, 226, 747, \dodoi{10.1093/mnras/226.4.747}

\bibitem[{{Just} {et~al.}(2019){Just}, {Kirby}, {Zaritsky}, {Rudnick}, {Desjardins}, {Cool}, {Moustakas}, {Clowe}, {De Lucia}, {Arag{\'o}n-Salamanca}, {Desai}, {Finn}, {Halliday}, {Jablonka}, {Mann}, {Poggianti}, {Bian}, \& {Liebst}}]{Jus19}
{Just}, D.~W., {Kirby}, M., {Zaritsky}, D., {et~al.} 2019, \apj, 885, 6, \dodoi{10.3847/1538-4357/ab44a0}

\bibitem[{{Kilborn} {et~al.}(2006){Kilborn}, {Forbes}, {Koribalski}, {Brough}, \& {Kern}}]{Kil06}
{Kilborn}, V.~A., {Forbes}, D.~A., {Koribalski}, B.~S., {Brough}, S., \& {Kern}, K. 2006, \mnras, 371, 739, \dodoi{10.1111/j.1365-2966.2006.10697.x}

\bibitem[{{Kleiner} {et~al.}(2019){Kleiner}, {Koribalski}, {Serra}, {Whiting}, {Westmeier}, {Wong}, {Kamphuis}, {Popping}, {Bekiaris}, {Elagali}, {For}, {Lee-Waddell}, {Madrid}, {Reynolds}, {Rhee}, {Shao}, {Staveley-Smith}, {Wang}, {Anderson}, {Collier}, {Ord}, \& {Voronkov}}]{Kle19}
{Kleiner}, D., {Koribalski}, B.~S., {Serra}, P., {et~al.} 2019, \mnras, 488, 5352, \dodoi{10.1093/mnras/stz2063}

\bibitem[{{Kleiner} {et~al.}(2021){Kleiner}, {Serra}, {Maccagni}, {Venhola}, {Morokuma-Matsui}, {Peletier}, {Iodice}, {Raj}, {de Blok}, {Comrie}, {J{\'o}zsa}, {Kamphuis}, {Loni}, {Loubser}, {Moln{\'a}r}, {Passmoor}, {Ramatsoku}, {Sivitilli}, {Smirnov}, {Thorat}, \& {Vitello}}]{Kle21}
{Kleiner}, D., {Serra}, P., {Maccagni}, F.~M., {et~al.} 2021, \aap, 648, A32, \dodoi{10.1051/0004-6361/202039898}

\bibitem[{{Koribalski} {et~al.}(2003){Koribalski}, {Gordon}, \& {Jones}}]{Kor03}
{Koribalski}, B., {Gordon}, S., \& {Jones}, K. 2003, \mnras, 339, 1203, \dodoi{10.1046/j.1365-8711.2003.06277.x}

\bibitem[{{Koribalski} {et~al.}(2018){Koribalski}, {Wang}, {Kamphuis}, {Westmeier}, {Staveley-Smith}, {Oh}, {L{\'o}pez-S{\'a}nchez}, {Wong}, {Ott}, {de Blok}, \& {Shao}}]{Kor18}
{Koribalski}, B.~S., {Wang}, J., {Kamphuis}, P., {et~al.} 2018, \mnras, 478, 1611, \dodoi{10.1093/mnras/sty479}

\bibitem[{{Koribalski} {et~al.}(2020){Koribalski}, {Staveley-Smith}, {Westmeier}, {Serra}, {Spekkens}, {Wong}, {Lee-Waddell}, {Lagos}, {Obreschkow}, {Ryan-Weber}, {Zwaan}, {Kilborn}, {Bekiaris}, {Bekki}, {Bigiel}, {Boselli}, {Bosma}, {Catinella}, {Chauhan}, {Cluver}, {Colless}, {Courtois}, {Crain}, {de Blok}, {D{\'e}nes}, {Duffy}, {Elagali}, {Fluke}, {For}, {Heald}, {Henning}, {Hess}, {Holwerda}, {Howlett}, {Jarrett}, {Jones}, {Jones}, {J{\'o}zsa}, {Jurek}, {J{\"u}tte}, {Kamphuis}, {Karachentsev}, {Kerp}, {Kleiner}, {Kraan-Korteweg}, {L{\'o}pez-S{\'a}nchez}, {Madrid}, {Meyer}, {Mould}, {Murugeshan}, {Norris}, {Oh}, {Oosterloo}, {Popping}, {Putman}, {Reynolds}, {Rhee}, {Robotham}, {Ryder}, {Schr{\"o}der}, {Shao}, {Stevens}, {Taylor}, {van{\^A} der Hulst}, {Verdes-Montenegro}, {Wakker}, {Wang}, {Whiting}, {Winkel}, \& {Wolf}}]{Kor20}
{Koribalski}, B.~S., {Staveley-Smith}, L., {Westmeier}, T., {et~al.} 2020, \apss, 365, 118, \dodoi{10.1007/s10509-020-03831-4}

\bibitem[{{Kormendy} \& {Kennicutt}(2004)}]{Kor04}
{Kormendy}, J., \& {Kennicutt}, Jr., R.~C. 2004, \araa, 42, 603, \dodoi{10.1146/annurev.astro.42.053102.134024}

\bibitem[{{Larson} {et~al.}(1980){Larson}, {Tinsley}, \& {Caldwell}}]{Lar80}
{Larson}, R.~B., {Tinsley}, B.~M., \& {Caldwell}, C.~N. 1980, \apj, 237, 692, \dodoi{10.1086/157917}

\bibitem[{{Lee-Waddell} {et~al.}(2019){Lee-Waddell}, {Koribalski}, {Westmeier}, {Elagali}, {For}, {Kleiner}, {Madrid}, {Popping}, {Reynolds}, {Rhee}, {Serra}, {Shao}, {Staveley-Smith}, {Wang}, {Whiting}, {Wong}, {Allison}, {Bhandari}, {Collier}, {Heald}, {Marvil}, \& {Ord}}]{LWa19}
{Lee-Waddell}, K., {Koribalski}, B.~S., {Westmeier}, T., {et~al.} 2019, \mnras, 487, 5248, \dodoi{10.1093/mnras/stz017}

\bibitem[{{Leroy} {et~al.}(2012){Leroy}, {Bigiel}, {de Blok}, {Boissier}, {Bolatto}, {Brinks}, {Madore}, {Munoz-Mateos}, {Murphy}, {Sandstrom}, {Schruba}, \& {Walter}}]{Ler12}
{Leroy}, A.~K., {Bigiel}, F., {de Blok}, W.~J.~G., {et~al.} 2012, \aj, 144, 3, \dodoi{10.1088/0004-6256/144/1/3}

\bibitem[{{Malphrus} {et~al.}(1997){Malphrus}, {Simpson}, {Gottesman}, \& {Hawarden}}]{Mal97}
{Malphrus}, B.~K., {Simpson}, C.~E., {Gottesman}, S.~T., \& {Hawarden}, T.~G. 1997, \aj, 114, 1427, \dodoi{10.1086/118574}

\bibitem[{{McMullin} {et~al.}(2007){McMullin}, {Waters}, {Schiebel}, {Young}, \& {Golap}}]{McM07}
{McMullin}, J.~P., {Waters}, B., {Schiebel}, D., {Young}, W., \& {Golap}, K. 2007, in Astronomical Society of the Pacific Conference Series, Vol. 376, Astronomical Data Analysis Software and Systems XVI, ed. R.~A. {Shaw}, F.~{Hill}, \& D.~J. {Bell}, 127

\bibitem[{{Michel-Dansac} {et~al.}(2010){Michel-Dansac}, {Duc}, {Bournaud}, {Cuillandre}, {Emsellem}, {Oosterloo}, {Morganti}, {Serra}, \& {Ibata}}]{Mic10}
{Michel-Dansac}, L., {Duc}, P.-A., {Bournaud}, F., {et~al.} 2010, \apjl, 717, L143, \dodoi{10.1088/2041-8205/717/2/L143}

\bibitem[{{Moore} {et~al.}(1996){Moore}, {Katz}, {Lake}, {Dressler}, \& {Oemler}}]{Moo96}
{Moore}, B., {Katz}, N., {Lake}, G., {Dressler}, A., \& {Oemler}, A. 1996, \nat, 379, 613, \dodoi{10.1038/379613a0}

\bibitem[{{Moran} {et~al.}(2006){Moran}, {Ellis}, {Treu}, {Salim}, {Rich}, {Smith}, \& {Kneib}}]{Mor06}
{Moran}, S.~M., {Ellis}, R.~S., {Treu}, T., {et~al.} 2006, \apjl, 641, L97, \dodoi{10.1086/504078}

\bibitem[{{Namumba} {et~al.}(2021){Namumba}, {Koribalski}, {J{\'o}zsa}, {Lee-Waddell}, {Jones}, {Carignan}, {Verdes-Montenegro}, {Ianjamasimanana}, {de Blok}, {Cluver}, {Garrido}, {S{\'a}nchez-Exp{\'o}sito}, {Ramaila}, {Thorat}, {Andati}, {Hugo}, {Kleiner}, {Kamphuis}, {Serra}, {Smirnov}, {Maccagni}, {Makhathini}, {Moln{\'a}r}, {Perkins}, {Ramatsoku}, {White}, \& {Loi}}]{Nam21}
{Namumba}, B., {Koribalski}, B.~S., {J{\'o}zsa}, G.~I.~G., {et~al.} 2021, \mnras, 505, 3795, \dodoi{10.1093/mnras/stab1524}

\bibitem[{{Nulsen}(1982)}]{Nul82}
{Nulsen}, P.~E.~J. 1982, \mnras, 198, 1007, \dodoi{10.1093/mnras/198.4.1007}

\bibitem[{{Oosterloo} {et~al.}(2010){Oosterloo}, {Morganti}, {Crocker}, {J{\"u}tte}, {Cappellari}, {de Zeeuw}, {Krajnovi{\'c}}, {McDermid}, {Kuntschner}, {Sarzi}, \& {Weijmans}}]{Oos10}
{Oosterloo}, T., {Morganti}, R., {Crocker}, A., {et~al.} 2010, \mnras, 409, 500, \dodoi{10.1111/j.1365-2966.2010.17351.x}

\bibitem[{{Oosterloo} {et~al.}(2018){Oosterloo}, {Zhang}, {Lucero}, \& {Carignan}}]{Oos18}
{Oosterloo}, T.~A., {Zhang}, M.~L., {Lucero}, D.~M., \& {Carignan}, C. 2018, arXiv e-prints, arXiv:1803.08263, \dodoi{10.48550/arXiv.1803.08263}

\bibitem[{{Pak} {et~al.}(2019){Pak}, {Lee}, {Jeong}, {Kim}, {Smith}, \& {Lee}}]{Pak19}
{Pak}, M., {Lee}, J.~H., {Jeong}, H., {et~al.} 2019, \apj, 880, 149, \dodoi{10.3847/1538-4357/ab2ad6}

\bibitem[{{Pak} {et~al.}(2021){Pak}, {Oh}, {Lee}, {Scott}, {Smith}, {van de Sande}, {Croom}, {D'Eugenio}, {Bekki}, {Brough}, {Foster}, {Barone}, {Kraljic}, {Jeong}, {Bland-Hawthorn}, {Bryant}, {Goodwin}, {Lawrence}, {Owers}, \& {Richards}}]{Pak21}
{Pak}, M., {Oh}, S., {Lee}, J.~H., {et~al.} 2021, \apj, 906, 43, \dodoi{10.3847/1538-4357/abc880}

\bibitem[{{Piffl} {et~al.}(2014){Piffl}, {Scannapieco}, {Binney}, {Steinmetz}, {Scholz}, {Williams}, {de Jong}, {Kordopatis}, {Matijevi{\v{c}}}, {Bienaym{\'e}}, {Bland-Hawthorn}, {Boeche}, {Freeman}, {Gibson}, {Gilmore}, {Grebel}, {Helmi}, {Munari}, {Navarro}, {Parker}, {Reid}, {Seabroke}, {Watson}, {Wyse}, \& {Zwitter}}]{Pif14}
{Piffl}, T., {Scannapieco}, C., {Binney}, J., {et~al.} 2014, \aap, 562, A91, \dodoi{10.1051/0004-6361/201322531}

\bibitem[{{Pisano} {et~al.}(2000){Pisano}, {Wilcots}, \& {Elmegreen}}]{Pis00}
{Pisano}, D.~J., {Wilcots}, E.~M., \& {Elmegreen}, B.~G. 2000, \aj, 120, 763, \dodoi{10.1086/301464}

\bibitem[{{Rasmussen} {et~al.}(2008){Rasmussen}, {Ponman}, {Verdes-Montenegro}, {Yun}, \& {Borthakur}}]{Ras08}
{Rasmussen}, J., {Ponman}, T.~J., {Verdes-Montenegro}, L., {Yun}, M.~S., \& {Borthakur}, S. 2008, \mnras, 388, 1245, \dodoi{10.1111/j.1365-2966.2008.13451.x}

\bibitem[{{Saponara} {et~al.}(2018){Saponara}, {Koribalski}, {Benaglia}, \& {Fern{\'a}ndez L{\'o}pez}}]{Sap18}
{Saponara}, J., {Koribalski}, B.~S., {Benaglia}, P., \& {Fern{\'a}ndez L{\'o}pez}, M. 2018, \mnras, 473, 3358, \dodoi{10.1093/mnras/stx2475}

\bibitem[{{Schneider} {et~al.}(1983){Schneider}, {Helou}, {Salpeter}, \& {Terzian}}]{Sch83}
{Schneider}, S.~E., {Helou}, G., {Salpeter}, E.~E., \& {Terzian}, Y. 1983, \apjl, 273, L1, \dodoi{10.1086/184118}

\bibitem[{{Scott} {et~al.}(2022){Scott}, {Cortese}, {Lagos}, {Brinks}, {Finoguenov}, \& {Coccato}}]{Sco22}
{Scott}, T.~C., {Cortese}, L., {Lagos}, P., {et~al.} 2022, \mnras, 511, 980, \dodoi{10.1093/mnras/stac118}

\bibitem[{{Serra} {et~al.}(2012){Serra}, {Oosterloo}, {Morganti}, {Alatalo}, {Blitz}, {Bois}, {Bournaud}, {Bureau}, {Cappellari}, {Crocker}, {Davies}, {Davis}, {de Zeeuw}, {Duc}, {Emsellem}, {Khochfar}, {Krajnovi{\'c}}, {Kuntschner}, {Lablanche}, {McDermid}, {Naab}, {Sarzi}, {Scott}, {Trager}, {Weijmans}, \& {Young}}]{Ser12}
{Serra}, P., {Oosterloo}, T., {Morganti}, R., {et~al.} 2012, \mnras, 422, 1835, \dodoi{10.1111/j.1365-2966.2012.20219.x}

\bibitem[{{Serra} {et~al.}(2013){Serra}, {Koribalski}, {Duc}, {Oosterloo}, {McDermid}, {Michel-Dansac}, {Emsellem}, {Cuillandre}, {Alatalo}, {Blitz}, {Bois}, {Bournaud}, {Bureau}, {Cappellari}, {Crocker}, {Davies}, {Davis}, {de Zeeuw}, {Khochfar}, {Krajnovi{\'c}}, {Kuntschner}, {Lablanche}, {Morganti}, {Naab}, {Sarzi}, {Scott}, {Weijmans}, \& {Young}}]{Ser13}
{Serra}, P., {Koribalski}, B., {Duc}, P.-A., {et~al.} 2013, \mnras, 428, 370, \dodoi{10.1093/mnras/sts033}

\bibitem[{{Serra} {et~al.}(2019){Serra}, {Maccagni}, {Kleiner}, {de Blok}, {van Gorkom}, {Hugo}, {Iodice}, {J{\'o}zsa}, {Kamphuis}, {Kraan-Korteweg}, {Loni}, {Makhathini}, {Moln{\'a}r}, {Oosterloo}, {Peletier}, {Ramaila}, {Ramatsoku}, {Smirnov}, {Smith}, {Spavone}, {Thorat}, {Trager}, \& {Venhola}}]{Ser19}
{Serra}, P., {Maccagni}, F.~M., {Kleiner}, D., {et~al.} 2019, \aap, 628, A122, \dodoi{10.1051/0004-6361/201936114}

\bibitem[{{Smith} {et~al.}(2007){Smith}, {Ruchti}, {Helmi}, {Wyse}, {Fulbright}, {Freeman}, {Navarro}, {Seabroke}, {Steinmetz}, {Williams}, {Bienaym{\'e}}, {Binney}, {Bland-Hawthorn}, {Dehnen}, {Gibson}, {Gilmore}, {Grebel}, {Munari}, {Parker}, {Scholz}, {Siebert}, {Watson}, \& {Zwitter}}]{Smi07}
{Smith}, M.~C., {Ruchti}, G.~R., {Helmi}, A., {et~al.} 2007, \mnras, 379, 755, \dodoi{10.1111/j.1365-2966.2007.11964.x}

\bibitem[{{Springob} {et~al.}(2005){Springob}, {Haynes}, {Giovanelli}, \& {Kent}}]{Spr05}
{Springob}, C.~M., {Haynes}, M.~P., {Giovanelli}, R., \& {Kent}, B.~R. 2005, \apjs, 160, 149, \dodoi{10.1086/431550}

\bibitem[{{Thilker} {et~al.}(2009){Thilker}, {Donovan}, {Schiminovich}, {Bianchi}, {Boissier}, {Gil de Paz}, {Madore}, {Martin}, \& {Seibert}}]{Thi09}
{Thilker}, D.~A., {Donovan}, J., {Schiminovich}, D., {et~al.} 2009, \nat, 457, 990, \dodoi{10.1038/nature07780}

\bibitem[{{Tully}(2015)}]{Tul15}
{Tully}, R.~B. 2015, \aj, 149, 171, \dodoi{10.1088/0004-6256/149/5/171}

\bibitem[{{van Gorkom} \& {Schiminovich}(1997)}]{vanG97}
{van Gorkom}, J., \& {Schiminovich}, D. 1997, in Astronomical Society of the Pacific Conference Series, Vol. 116, The Nature of Elliptical Galaxies; 2nd Stromlo Symposium, ed. M.~{Arnaboldi}, G.~S. {Da Costa}, \& P.~{Saha}, 310

\bibitem[{{Yun} {et~al.}(1994){Yun}, {Ho}, \& {Lo}}]{Yun94}
{Yun}, M.~S., {Ho}, P.~T.~P., \& {Lo}, K.~Y. 1994, \nat, 372, 530, \dodoi{10.1038/372530a0}

\bibitem[{{Zabludoff} \& {Mulchaey}(1998)}]{Zab98}
{Zabludoff}, A.~I., \& {Mulchaey}, J.~S. 1998, \apj, 496, 39, \dodoi{10.1086/305355}

\end{thebibliography}

          
\facilities{JVLA, BOAO:1.8m, GALEX, DES, WISE}
\software{Astronomical Image Processing System (AIPS; \citealt{Gre03}),  
          Common Astronomy Software Applications (CASA; \citealt{McM07}), 
          Cube Analysis and Rendering Tool for Astronomy (CARTA; \citealt{Com21}),
          Image Reduction an Analysis Facility (IRAF; \citealt{Jed87})         }





\end{document}